\documentclass[prb,twocolumn,showpacs,superscriptaddress]{revtex4-1}

\usepackage{graphics}
\usepackage{amsmath,amsfonts}
\usepackage{epsf}
\usepackage{amssymb}
\usepackage{graphicx}
\newcommand{\ket}[1]{ |#1 \rangle}
\newcommand{\bra}[1]{ \langle #1|}

\newcommand{\kv}{\mathbf{k}}

\newcommand{\qv}{\mathbf{q}}

\newcommand{\rv}{\mathbf{r}}

\newcommand{\Rv}{\mathbf{R}}
\newcommand{\vu}{\mathnormal{v}}
\newcommand{\cd}{\cdot}
\newcommand{\brho}{\mbox{\boldmath$\rho$}}
\newcommand{\eps}{\epsilon}

\begin{document}

\title{Performance of local orbital basis sets
in the self-consistent Sternheimer method for dielectric matrices
of extended systems}

\author{Hannes H\"ubener}
\affiliation{Department of Materials, University of Oxford, Oxford OX1 3PH,
United Kingdom}

\author{Miguel A. \surname{P\'erez-Osorio}} 
\affiliation{Department of Materials, University of Oxford, Oxford OX1 3PH,
United Kingdom}
\affiliation{Centre d'Investigaci\'o en Nanoci\`encia i Nanotecnologia-CIN2
(CSIC-ICN), Campus UAB, Bellaterra, Spain}

\author{Pablo Ordej\'on}
\affiliation{Centre d'Investigaci\'o en Nanoci\`encia i Nanotecnologia-CIN2
(CSIC-ICN), Campus UAB, Bellaterra, Spain}

\author{Feliciano Giustino}
\affiliation{Department of Materials, University of Oxford, Oxford OX1 3PH,
United Kingdom}

\begin{abstract}
We present a systematic study of the performance of numerical pseudo-atomic
orbital basis sets
in the calculation of dielectric matrices of extended systems using 
the self-consistent Sternheimer approach of 
[F. Giustino \textit{et al.}, Phys. Rev. B \textbf{81}(11), 115105 (2010)]. 
In order to cover a range of systems, from more insulating to more metallic
character, we discuss results for the three semiconductors diamond, silicon, 
and germanium. Dielectric matrices calculated using our method
fall within 1-3\% of reference planewaves calculations, demonstrating that 
this method is promising. We find that polarization orbitals are critical 
for achieving good agreement with planewaves calculations,
and that only a few additional $\zeta$'s are required for obtaining
converged results, provided the split norm is properly optimized.
Our present work establishes the validity of local orbital basis sets
and the self-consistent Sternheimer approach for the calculation of
dielectric matrices in extended systems, and prepares the ground for 
future studies of electronic excitations using these methods.
\end{abstract} 

\date{\today{}}

\maketitle

\section{Introduction}

In recent years electronic structure codes based on local orbital basis sets 
have proven successful in describing complex 
systems involving several thousands of atoms.\cite{soler02,skylaris05,bowler10}
The key concept behind the use of local orbital basis sets in the solid state 
is that the ground-state electronic density matrix is exponentially localized 
in insulators.\cite{kohn96} As a consequence of this localization, the
representation 
of one-particle operators in a local orbital basis leads to strictly sparse 
matrices. It is therefore possible to solve the electronic structure problem,
e.g.\ the Kohn-Sham equations of density-functional theory (DFT), using 
numerical methods whose complexity scales linearly as a function of system 
size.\cite{ordejon93,mauri93,ordejon96,goedecker99}

A natural question arising is whether such local orbital basis sets would
also be advantageous in the study of electronic excitations. Since these basis 
sets are optimized for providing an accurate description of ground-state properties, 
it is not clear {\it a priori} how to exploit them for calculating excited 
state properties. For example, in the specific case of $GW$ 
calculations\cite{hedin65,hybertsen86}
the Green's function and the screened Coulomb interaction are both evaluated by using
expansions over unoccupied electronic states. However, local orbital basis sets 
typically provide a poor description of unoccupied states, and the 
convergence as a function of basis size is not systematic as in the case of  
planewaves basis sets\cite{ihm79} or finite difference 
methods.\cite{chelikowsky94} In this context it would be desirable to develop
new methods 
for electronic excitations which (i) retain the favorable scaling of local
orbital 
basis sets, and (ii) do not require the explicit calculation of unoccupied states.

Recently several schemes have been proposed in order to reduce or avoid
the evaluation of unoccupied states in excited-state 
calculations.\cite{bruneval08,berger10,wilson08,wilson09,giustino10,umari10}
In Refs.~\onlinecite{bruneval08,berger10}
a small number of unoccupied states or even one single state are used to
effectively replace the expansion over the conduction manifold.
The authors of Ref.~\onlinecite{wilson08,wilson09} use selected eigenvectors
of the dielectric matrix in order to evaluate the screened Coulomb interaction
without explicitly performing sums over empty states.
In Ref.~\onlinecite{umari10} a Lanczos algorithm is used in order to calculate
the dielectric matrix using continued fractions, without performing sums 
over empty states. 

We here consider the scheme proposed in Ref.~\onlinecite{giustino10}
for evaluating the screened Coulomb interaction in extended systems.
In this scheme the expansion over unoccupied states and the
matrix inversion are avoided altogether, and replaced by the self-consistent
solution of Sternheimer equations.
In a recent work\cite{huebener12} we have extended the scheme of 
Ref.~\onlinecite{giustino10} to the case of local orbital basis sets,
and implemented our new scheme using the {\tt SIESTA} package.\cite{soler02}
Ref.~\onlinecite{huebener12} represents, to our knowledge, the first
implementation
of the
self-consistent Sternheimer equation in a local orbitals basis for extended systems.

In the present manuscript we complement and extend the work of
Ref.~\onlinecite{huebener12}
by systematically comparing the performance of numerical pseudo-atomic orbital
basis sets
with standard plane-waves results. Since most excited-state calculations
involve kernels describing the nonlocal and time-dependent dielectric screening 
of the material, we here focus on the the frequency- and wavevector- dependent 
dielectric matrix of representative semiconductors.

The manuscript is organized as follows. In Sec.~\ref{sec:method} we describe our
methodology. In particular, in Sec.~\ref{sec:method-0} we summarize the self-consistent 
Sternheimer scheme for the screened Coulomb interaction and the inverse dielectric
matrix. In Secs.~\ref{sec:method-1} and \ref{sec:method-2} we specialize the formalism 
to the case of local orbital basis sets and periodic systems, respectively.
In Sec.~\ref{sec:comput-details} we briefly describe 
the basis set of {\tt SIESTA} that we use in all our calculations
(Sec.~\ref{sec:method-3}) and we provide details on the structural and
convergence 
parameters used in the calculations (Sec.~\ref{sec:method-4}).
In Sec.~\ref{sec:results} we discuss our results, focusing on the three semiconductors
diamond, silicon, and germanium. This choice allows us to understand how the performance
of our methodology varies when moving from more insulating (diamond) to
more metallic systems (germanium). In particular we investigate both the static
dielectric matrices (Sec.~\ref{sec:static}), and the frequency- and wavevector-dependent
dielectric functions (Sec.~\ref{sec:frequency}).
In Sec.~\ref{sec:conclusions} we summarize our findings and draw our conclusions.
We leave to Appendix~\ref{app:qR} some technical details of our implementation.

\section{Theoretical methodology}\label{sec:method}

\subsection{Self-consistent Sternheimer approach for the screened Coulomb interaction}
\label{sec:method-0}

The first order change of the valence Kohn-Sham wavefunction $\psi_\vu$ due 
to the perturbation $\Delta V$ is expressed by the Sternheimer equation as:
  \begin{equation}\label{eq.dfpt1}
  (\hat{H}-\epsilon_\vu)\Delta \psi_\vu = -(1-\hat{P}_v)\Delta V \psi_\vu,
  \end{equation}
where $\hat{H}$ is the Kohn-Sham Hamiltonian and 
$1-\hat{P}_v$ is the projector on the conduction manifold. 
The corresponding change in the valence electron density is given by:
  \begin{equation}\label{eq.dfpt2}
  \Delta n = 4 {\rm Re} \sum_\vu \psi_\vu^*\Delta \psi_\vu,
  \end{equation}
where the prefactor also takes into account the spin degeneracy (we refer
to spin-unpolarized systems for simplicity).
The density variation of Eq.~(\ref{eq.dfpt2}) induces a change in the
self-consistent potential $\Delta V$ experienced by the electrons (through the
Hartree and exchange-correlation terms), therefore Eqs.~(\ref{eq.dfpt1}) and (\ref{eq.dfpt2})
must be solved iteratively. This procedure is at the core of density-functional 
perturbation theory for lattice dynamics.\cite{baroni87,baroni01}

In the following we summarize the extension of the self-consistent Sternheimer
approach to the calculation of the screened Coulomb interaction $W$ and the inverse 
dielectric matrix $\eps^{-1}$, as derived in Ref.~\onlinecite{giustino10}.
These quantities are both functions
of the real-space variables $\rv$, $\rv'$, and of the frequency $\omega$,
and are related to the bare Coulomb interaction $v$ through:
  \begin{equation}
  W(\rv,\rv';\omega) = \int d\rv'' \eps^{-1} (\rv,\rv'';\omega) v(\rv'',\rv').
  \end{equation}
The calculation of $\eps^{-1}$ starting from $W$ is straightforward in reciprocal space,
therefore in the following we focus on the screened Coulomb interaction.
We first parametrize the frequency and one spatial variable of $W$ by
defining the potential $\Delta V_{[\rv,\omega]}(\rv') = W(\rv,\rv';\omega)$.
This potential induces a change $\Delta \psi^\pm_{\vu[\rv,\omega]} $ 
of the valence wavefunction $\psi_\vu$ given by: 
  \begin{equation}\label{eq:stern}     
  (\hat{H}-\eps_\vu \pm \omega)\Delta \psi^\pm_{\vu[\rv,\omega]} = 
  -(1-\hat{P}_{\vu})\Delta V_{[\rv,\omega]}\psi_\vu .
  \end{equation}
The associated variation of the density matrix is
  \begin{equation}
  \Delta n_{[\rv,\omega]} =
  2 \sum_{\vu,\sigma=\pm}\psi^*_\vu \Delta\psi^\sigma_{\vu[\rv,\omega]}.
  \end{equation}
In the random-phase approximation (RPA) this variation generates the screening
Hartree potential \cite{giustino10}
  \begin{equation}\label{eq:hartree}
  \Delta V^H_{[\rv,\omega]}(\rv') =
  \int d\rv''\Delta n_{[\rv,\omega]}(\rv'')\vu(\rv'',\rv'),
  \end{equation}
which is added to the bare Coulomb potential in order to obtain the total self-consistent 
potential appearing in Eq.~(\ref{eq:stern}):
  \begin{equation}\label{eq:indpot}
  \Delta V_{[\rv,\omega]}(\rv')=  v(\rv,\rv') + \Delta V^H_{[\rv,\omega]}(\rv').
  \end{equation}
The iterative self-consistent solution of Eqs.~(\ref{eq:stern})-(\ref{eq:indpot})
for all the values of the parameters $\rv$ and $\omega$ yields the screened Coulomb 
interaction $W$.
This scheme is exactly equivalent\cite{giustino10} 
to evaluating $W$ via Hedin's equation
  \begin{equation}
  W = \vu + WP\vu,
  \end{equation}
$P$ being the RPA polarizability.

\subsection{Self-consistent Sternheimer approach with local orbitals}\label{sec:method-1}

In order to implement the method described in Sec.~\ref{sec:method-0} within
the basis of local orbitals $\phi_i(\rv')$ we expand the wavefunctions and their 
first order variations as:
  \begin{eqnarray}\label{eq:expVal}
  \psi_\vu(\rv') &=& \sum_i c_{\vu i} \phi_i(\rv'),\\
  \Delta\psi_{\vu[\rv,\omega]}(\rv') &=& 
  \sum_i \Delta c_{\vu i[\rv,\omega]} \phi_i(\rv') \label{def:dc} .
  \end{eqnarray}
The expansions would be exact if the local orbitals were to span the
entire Hilbert space of the single-particle Hamiltonian. In practice
the representation of the valence wavefunctions is expected to be accurate,
while the description of the variations requires some care because
they arise from the conduction manifold.
By replacing Eqs.~(\ref{eq:expVal})-(\ref{def:dc}) 
inside Eq.~(\ref{eq:stern}) and performing 
scalar products of both sides with basis functions we obtain the matrix equations:
  \begin{equation}\label{eq:sternMatrix}
  \left[\mathbf{H} - (\epsilon_\vu\pm \omega)\mathbf{S}\right]
  \Delta \mathbf{c}^\pm_{\vu[\rv,\omega]} =
  -\left[\mathbf{1} -\mathbf{S}\brho^T \right]
  \Delta \mathbf{V}_{[\rv,\omega]} \mathbf{c}_{\vu},
  \end{equation}
where the Hamiltonian $\mathbf{H}$, overlap $\mathbf{S}$, density $\brho$,
and perturbation $\Delta \mathbf{V}_{[\rv,\omega]}$ matrices are defined 
in the usual notation as:
  \begin{eqnarray}
  H_{ij} &=& \bra{\phi_i} H \ket{\phi_j},\\
  S_{ij} &=& \bra{\phi_i} \phi_j\rangle,\\
  \rho_{ij} &=& \sum_{v'} c^*_{v'i}c_{v'j},\\
  \Delta V_{ij[\rv,\omega]} & = & \bra{\phi_i} \Delta V_{[\rv,\omega]} 
  \ket{\phi_j} \label{eq:dV}.
  \end{eqnarray}
The Hamiltonian, the overlap, and the density matrices are readily available 
in any local orbital DFT implementation. The matrix elements $\Delta V_{ij[\rv,\omega]}$ 
require careful consideration. 
The iterative self-consistent solution of the Sternheimer equation starts
with $\Delta V_{[\rv,\omega]}(\rv')$ initialized to the bare Coulomb potential 
$v(\rv,\rv')$. This is a nonlocal potential, therefore an expansion in the
local orbital basis would require the product functions $\phi_i(\rv)
\phi_j(\rv')$.\cite{aryasetiawan94} Using such
expansion\cite{rohlfing95,blase04,koval10,foerster11,blase11} in
Eq.~(\ref{eq:dV}) would ultimately 
lead to the evaluation of four-point integrals, and would also introduce
three-point overlaps in the Sternheimer
equations. In addition, the validity
of such expansion for the Coulomb potential is not guaranteed.
In order to avoid this difficulty we evaluate
the integral of Eq.~(\ref{eq:dV}) in real space. This real-space  
integral is part of the {\tt SIESTA} implementation.
We stress that the evaluation of Eq.~(\ref{eq:dV}) in real space does not suffer from 
representability issues associated with a product-basis expansion,
and we do not make use of the expansion of the identity operator in the local
orbital basis.

The solution of Eq.~(\ref{eq:sternMatrix}) yields the vector $\Delta \bf{c}$
of the linear response coefficients which are used to construct the
variation of the density matrix:
  \begin{equation}\label{eq:dn}
  \Delta n_{ij[\rv,\omega]} = 2\sum_{\vu,\sigma=\pm} c^*_{\vu i}
  \Delta c^\sigma_{\vu j[\rv,\omega]}.
  \end{equation}
For the evaluation of the induced Hartree potential Eq.~(\ref{eq:hartree}),
first the variation of the density matrix is calculated on the real-space grid:
  \begin{equation}
  \Delta n_{[\rv,\omega]}(\rv') = 
  \sum_{ij} \Delta n_{ij[\rv,\omega]} \phi^*_i(\rv')\phi^*_j(\rv'),
  \end{equation}
then standard Fourier-transform techniques are employed.
The convergence of the scf-cycle is tested on the variation of the density matrix 
Eq.~(\ref{eq:dn}).

The self-consistent procedure is carried out independently 
for the parameters $\rv$ and $\omega$. Apart from leading to a trivial
parallelization, this scheme has the advantage that the parametrized
space variable $\rv$ can be represented on a coarser grid than the one 
used for $\rv'$ and the real-space integrals. 
Indeed, the accuracy required for operations in $\rv'$ is the same
needed to describe the Kohn-Sham Hamiltonian, while the accuracy for 
$\rv$ is the one needed to represent the screened Coulomb interaction.

\subsection{Self-consistent Sternheimer approach with local orbitals
for periodic systems}\label{sec:method-2}

In order to address extended periodic systems and compare our results
with calculations based on planewaves basis sets, in this section we
specialize the formalism of Sec.~\ref{sec:method-1} to the case of crystalline
solids. We start by introducing a new basis set which satisfies Bloch's theorem
from the outset. For this purpose we define the cell-periodic functions:
  \begin{equation}\label{eq.blochbasis}
  \phi_{i\kv}(\rv') = \sum_{\Rv}e^{-i\kv\cd(\rv'-\Rv)}\phi_{i}(\rv'-\Rv),
  \end{equation}
where $\kv$ is a point in the Brillouin zone and $\Rv$ a lattice vector.
In Eq.~(\ref{eq.blochbasis}) the orbitals $\phi_{i}(\rv')$ are the same
as in Eq.~(\ref{eq:expVal}), except that here they only span the unit cell of the
crystal. It is immediate to verify that the functions $e^{i\kv\cd\rv'}\phi_{i\kv}(\rv')$
satisfy Bloch's theorem, and that the basis functions $\phi_{i\kv}$ are periodic.

In analogy with Eqs.~(\ref{eq:expVal}), (\ref{def:dc}) we expand the periodic
part $u_{\vu\kv}$ of the Kohn-Sham eigenfunctions 
$\psi_{\vu\kv}(\rv') = e^{i\kv\cd\rv'}u_{\vu\kv}(\rv')$ using the basis $\phi_{i\kv}$:
  \begin{eqnarray}
  u_{v\kv}(\rv') &=& \sum_i c_{vi\kv}\phi_{i\kv}(\rv'),\\
  \Delta u_{v\kv[\qv,\rv]}(\rv') &=&\sum_i \Delta c_{vi\kv[\qv,\rv]}\phi_{i\kv+\qv}(\rv').
  \end{eqnarray}
By rewriting Eq.~(\ref{eq:stern}) for a periodic system\cite{giustino10}
and taking scalar products with basis functions 
we obtain the analogue of Eq.~(\ref{eq:sternMatrix}) for crystalline systems:
  \begin{equation}\label{eq:ksternMatrix}
  \begin{split}
  &\left[\mathbf{H}_{\kv+\qv} - (\epsilon_{\vu\kv}\pm \omega)\mathbf{S}_{\kv+\qv}\right]
  \Delta \mathbf{c}^\pm_{\vu \kv[\qv,\rv,\omega]} = \\
  & \qquad\qquad\; -\left[\mathbf{1} - \mathbf{S}_{\kv+\qv}\brho^T_{\kv+\qv}\right]
  \Delta \mathbf{V}_{\kv[\qv,\rv,\omega]} \mathbf{c}_{\vu\kv}.
  \end{split}
  \end{equation}
In this case the Hamiltonian and the overlap matrices are defined as:
  \begin{eqnarray}
  H_{ij\kv} &= &
   \sum_{\Rv}e^{i\kv\cd\Rv}\int d\rv' \phi_{i}(\rv')H(\rv')\phi_{j}(\rv'-\Rv),
  \label{eq.Ham.k}\\
  S_{ij\kv} &=& 
   \sum_{\Rv}e^{i\kv\cd\Rv}\int d\rv' \phi_{i}(\rv') \phi_{j}(\rv'-\Rv) .
  \end{eqnarray}
The perturbation matrix reads:
  \begin{eqnarray}
  && \hspace{-1.0cm}
  \Delta V_{ij\kv[\qv,\rv,\omega]} =
  \int d\rv' \phi^*_{i\kv+\qv}(\rv') \Delta v_{[\qv,\rv]}(\rv') \phi_{j\kv}(\rv') 
  \nonumber \\
  & &  \hspace{-0.9cm} = 
  \sum_{\Rv}e^{i\kv\cd\Rv}\int d\rv' \phi_{i}(\rv')\Delta v_{[\qv,\rv]}(\rv')
  e^{i\qv\cd\rv'}\phi_{j}(\rv'-\Rv),\label{eq.dVmatrix}
  \end{eqnarray}
where $\Delta v_{[\qv,\rv,\omega]}(\rv')$ is the cell-periodic component of the
perturbation with wavevector $\qv$:\cite{giustino10}
  \begin{equation}
  \Delta V_{[\rv,\omega]}(\rv')  =  \frac{1}{N_{\qv}}\sum_{\qv}e^{i\qv\cd(\rv'-\rv)}
  \Delta v_{[\qv,\rv,\omega]}(\rv').
  \end{equation}
In Eqs.~(\ref{eq.Ham.k})-(\ref{eq.dVmatrix}) the integrals extend over the unit cell.
This implies that the sums on lattice vectors $\Rv$ effectively include
the unit cells containing basis orbitals which have a non-vanishing contribution
in the fundamental unit cell. This is consistent with the standard procedure
for calculating matrix elements in the {\tt SIESTA} code.\cite{soler02}

For the calculation of the self-consistent potential 
in Eq.~(\ref{eq:ksternMatrix})
we analyze the density variation in Bloch components as:
  \begin{equation}
  \Delta n_{[\rv,\omega]}(\rv') = \frac{1}{N_{\qv}}
  \sum_\qv e^{i\qv\cd\rv'} \Delta n_{[\qv,\rv,\omega]}(\rv'),
  \end{equation}
and calculate the periodic part $\Delta n_{[\qv,\rv,\omega]}(\rv')$ on the
real space grid using:
  \begin{eqnarray}\nonumber
  & & \hspace{-1.0cm}\Delta n_{[\qv,\rv,\omega]}(\rv') =
  \frac{2}{N_{\kv}}\sum_{v\kv\sigma=\pm}\sum_{ ij}c^*_{vi\kv}
  \Delta c^\sigma_{vj\kv[\qv,\rv,\omega]} \\
  &  &\hspace{-.8cm} \times\sum_{\Rv\Rv'}e^{-i\kv\cd(\Rv-\Rv')}e^{i\qv\cd\Rv'}
  e^{-i\qv\cd\rv'}\phi_{i}(\rv'-\Rv)\phi_{j}(\rv'-\Rv') \label{eq:dn_matrix}.
  \end{eqnarray}
In the previous two equations $N_{\kv}$ and $N_{\qv}$ are the number of
wavevectors used to sample the Brillouin zone (we assume uniform sampling
only for simplicity of notation).

We point out that the presence of finite $\qv$-vectors in the above equations 
introduces non-trivial terms in the expressions for the variation of the
charge density and the associated induced potential. Loosely speaking,
the phase factors $\exp(i\qv\cd\rv)$ are calculated on the real-space grid,
while the phase factors $\exp(i\qv\cd\Rv)$ are added at the level of the
matrix elements. These terms require some care in a practical implementation,
as described in detail in Appendix~\ref{app:qR}.

\section{Computational details}\label{sec:comput-details}

\subsection{Local orbital basis}\label{sec:method-3}

The method presented in Sec.~\ref{sec:method} has been implemented
using the {\tt SIESTA} code as the starting software platform.
The basis functions $\phi_i$ are numerical pseudo-atomic
orbitals.\cite{soler02} For each orbital, the radial part is obtained by
solving
the radial Schr\"odinger equation for a pseudo-atom based on Troullier-Martins
psedupotentials.\cite{troullier91}
The basis orbitals are strictly localized within a preset cutoff radius,
which is controlled by a so-called ``energy shift'' parameter.\cite{artacho99}
This parameter is uniquely defined for a given calculation.
The basis of numerical pseudo-atomic orbitals can be augmented by
associating multiple radial functions with the same principal 
atomic quantum number. Such additional functions are constructed in {\tt SIESTA}
using the split-norm procedure,\cite{artacho99} and are denoted
in the quantum chemistry literature as ``multiple-$\zeta$''.
The split-norm construction guarantees that the additional $\zeta$'s
exhibit a smaller cutoff radius w.r.t. the originating radial function.
This feature ensures that the spatial extent of the basis orbitals
is dictated by the energy-shift parameter, regardless of how many additional
$\zeta$'s are used.
Additional flexibility in the basis set is usually achieved by augmenting
this basis through ``polarization orbitals''. Such orbitals are obtained
by solving the Schr\"odinger equation for the outermost shell of the 
pseudo-atom in the presence of a small electric
field.\cite{artacho99,portal96}

These polarization orbitals are generally needed for high accuracy and it
has been found that for most ground-state quantities of interest
double-$\zeta$ basis sets including polarization orbitals (DZP) yield
results in good agreement with standard planewaves
calculations.\cite{junquera01}

\subsection{Structural and convergence parameters}\label{sec:method-4}

For diamond, silicon, and germanium we
calculate the inverse dielectric matrix $\epsilon^{-1}_{{\bf G}{\bf G}'}(\qv,\omega)$ 
using the method described in Sec.~\ref{sec:method}
with the initial perturbation in Eq.~(\ref{eq:stern}) set to 
$\Delta V_{[\rv,\omega]}(\rv')={\rm exp}[i(\qv+{\bf G})\cdot\rv']$, 
and by taking the Fourier component of the resulting self-consistent
potential corresponding to the wavevector ${\bf G}'$.
The dielectric function is obtained as 
$\epsilon(\qv,\omega) = 1/\epsilon^{-1}_{{\bf 0}{\bf 0}}(\qv,\omega)$
and the macroscopic dielectric constant as $\epsilon_0=\epsilon(\qv \rightarrow
0,\omega=0)$.

We perform calculations using the local-density
approximation\cite{ceperley80,perdew81} to density-functional theory. Only
valence electrons
are described, and the core-valence interaction is taken into account through
norm-conserving pseudopotentials.\cite{troullier91}
In the following we test the method described in Sec.~\ref{sec:method} 
by considering a range of possibilities for the local orbitals basis sets.
The lattice parameters are set to 6.74 au, 10.26 au, and 10.68 au for
diamond, silicon, and germanium, respectively.\cite{hybertsen87}
The dielectric matrices are calculated by sampling the Brillouin zone
with a shifted $10\times10\times10$ mesh in the case of diamond and silicon,
and a $12\times12\times12$ mesh for germanium. When using a triple-$\zeta$
polarized (TZP) basis we obtain
the direct band gaps 5.60 eV, 2.55 eV, and 0.04 eV for diamond, silicon,
and germanium, respectively, in line with standard planewaves calculations.

We use the energy-shift parameter of 10 meV for all three materials. These parameters lead to localization
radii of 7.4 \AA, 9.3 \AA, and 9.5 \AA\  for diamond, silicon and germanium,
respectively, that are larger
than those adopted in standard ground-state calculations using {\tt SIESTA}. 
Results for standard values of the energy-shift parameters and localization 
radii are reported in Ref.~\onlinecite{huebener12}.

\begin{figure}
  \resizebox{0.9\columnwidth}{!}{\includegraphics{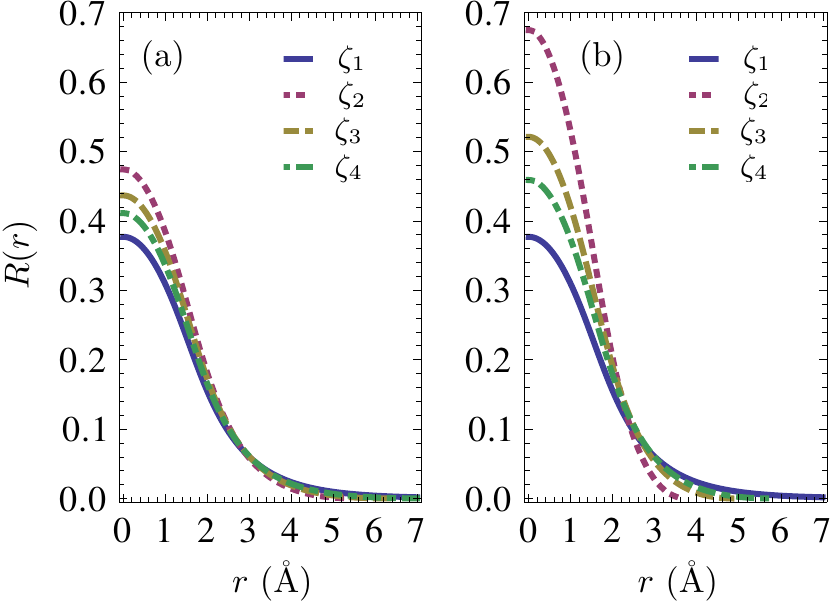}}
  \caption{
  Radial component of the Si-$2p$ pseudo-atomic orbitals corresponding to the
  4 $\zeta$ basis (4Z). \textbf{(a)} Multiple $\zeta$'s generated using a split norm
  of 0.15. In this case the additional basis functions are very similar to
  the first and second $\zeta$ and it is not advantageous to increase 
  the basis size. \textbf{(b)} Multiple $\zeta$'s generated using a split 
  norm of 0.5. In these case the additional orbitals are significantly different
  from the first and second $\zeta$ and we can expect that an increase of the
  basis size will improve the accuracy of the calculation.
  \label{fig:zetas}}
  \end{figure}

By tuning the energy-shift parameter and the split norm it is possible
to generate multiple $\zeta$ orbitals with a varying degree of localization. 
For example, in the case of silicon, if we use a split norm of 0.15, we obtain 
a localization radius of 9.3 \AA\ for the first $\zeta$
corresponding to the Si-$2p$ orbital, and a radius of 5.5 \AA\ for the second $\zeta$.
Additional $\zeta$ functions have by construction localization radii
between those of the first and of the second $\zeta$ (Fig.~\ref{fig:zetas}).
A small value of the split norm leads to $\zeta$ functions with very similar
localization radii and shape [Fig.~\ref{fig:zetas}(a)]. Larger values of the
split norm lead to a more even distribution of radii and allow for more
flexibility in the multiple-$\zeta$ basis [Fig.~\ref{fig:zetas}(b)].
We performed calculations for several values of the split norm between 0.15
and 0.5, and in Sec.~\ref{sec:static} discuss our results for the two
ends of this range. The standard value of the split norm
in {\tt SIESTA} calculations is 0.15.

For comparison we also perform standard planewaves calculations using
the {\tt ABINIT} package\cite{abinit} and the {\tt YAMBO} code\cite{yambo},
using the same Brillouin-zone grids and pseudopotentials\footnote{ We note that
there is a small difference between {\tt SIESTA} and {\tt ABINIT} in how the
local part of the pseudopotential is constructed.}.
In the planewaves calculations dielectric matrices are obtained within
the random-phase approximation using the Adler-Wiser
formulation.\cite{Adler,Wiser}
In all cases the calculations are found to be converged by using 92 
unoccupied electronic states. 
We use planewaves cutoffs of 20 Ry for silicon and germanium and 60 Ry for
diamond for the ground-state calculations. The corresponding planewaves cutoffs
for the dielectric
matrices are 12 Ry, 6.9 Ry, and 6.9 Ry for diamond, silicon, and germanium,
respectively.

While we carefully set all the parameters of the plane-waves calculations 
in order to make the comparison as accurate as possible, there 
remains one systematic difference in how the long-wavelength limit $\qv \rightarrow 0$ 
is taken in the calculation of the dielectric constant. 
We obtain this limit by considering a small but finite wavevector 
($q =0.01\,2\pi/a$, $a$ being the lattice parameter), while the {\tt YAMBO} 
code calculates this limit analytically. 
The two treatments are in principle equivalent, but we cannot rule out 
that this difference might result in small differences between 
the results presented in the following section.

\section{Results and discussion}\label{sec:results}

The purpose of this section is to study the convergence of the calculated
dielectric matrices with the size and type of the local orbital basis,
and to perform a systematic comparison with reference planewaves calculations.
We discuss our results for the dielectric matrices of
diamond, silicon, and germanium. We start with silicon since this has
been the benchmark semiconductor in a number of previous studies of
dielectric screening and quasiparticle methods.

\subsection{Macroscopic dielectric constants}\label{sec:static}

\subsubsection{Silicon}\label{sec.silicon}

\begin{figure}
  \begin{center}
  \resizebox{0.9\columnwidth}{!}{\includegraphics{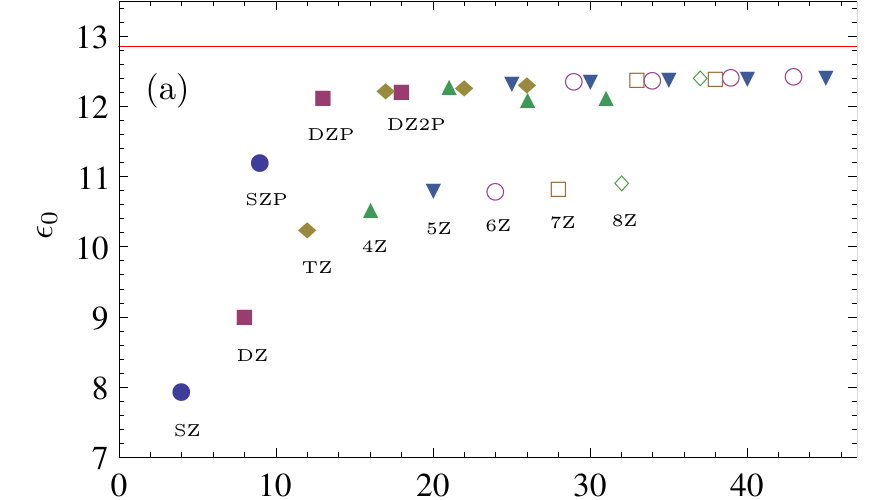}}
  \resizebox{0.9\columnwidth}{!}{\includegraphics{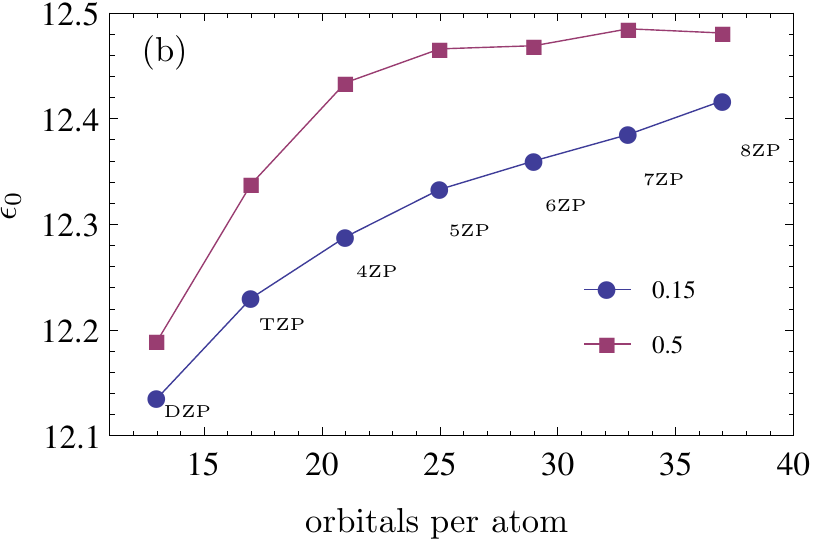}}
  \end{center}
  \caption{
  \textbf{(a)}: 
  Calculated macroscopic dielectric constant of silicon as a function
  of basis size, given in terms of orbitals per atom.
  The reference planewaves calculation is indicated by the horizontal red line.
  The datapoints cluster around two distinct curves: the upper curve
  corresponds to polarized basis sets, the lower curve to unpolarized basis sets.
  The number of $\zeta$ functions included is indicated by the labels SZ,
  DZ, TZ etc. The number of polarization orbitals for a given 
  number of $\zeta$'s increases towards the right-hand side, as indicated
  for the case of the DZ basis. The energy shift is 10 meV and the split norm is 0.15.
  \textbf{(b)}: 
  Calculated macroscopic dielectric constant of silicon as a function 
  of basis size, for two different values of the split norm.
  \label{fig:si_static}}
  \end{figure}

Figure \ref{fig:si_static}(a) shows the calculated macroscopic dielectric
constant of silicon as a function of basis size, for a split norm of 0.15. 
The smallest possible
basis includes 4 orbitals per atoms (1 $s$ orbital and 3 $p$ orbitals)
and is referred to as the single-$\zeta$ basis (SZ).
The basis sets with and without polarization orbitals appear 
to converge to different asymptotic values. The plateau of the
polarized basis set is $\eps_{0}=12.41$ and is reached with the TZP
basis (17 orbitals per atom). This value is within 3\% of our reference
planewaves result $\eps_{0}=12.85$.
Interestingly the minimal polarized basis (SZP, 9~orbitals per atom)
yields values which are within 10\% of the reference planewaves result.

Figure \ref{fig:si_static}(b) shows the effect of the split norm on the
convergence of the macroscopic dielectric constant as a function of basis size.
We observe that by increasing the split norm the calculated dielectric
constant converges to the planewaves value more rapidly. We assign this
trend to the fact that a larger split norm leads to a wider range of localization
radii spanned by the additional basis functions, and hence improves the
completeness of the basis set.

\subsubsection{Diamond}

\begin{figure}
  \resizebox{0.9\columnwidth}{!}{\includegraphics{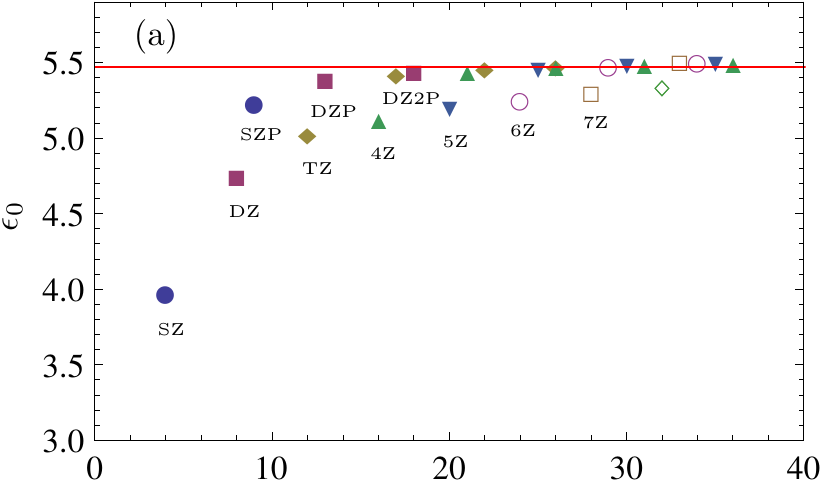}}
  \resizebox{0.9\columnwidth}{!}{\includegraphics{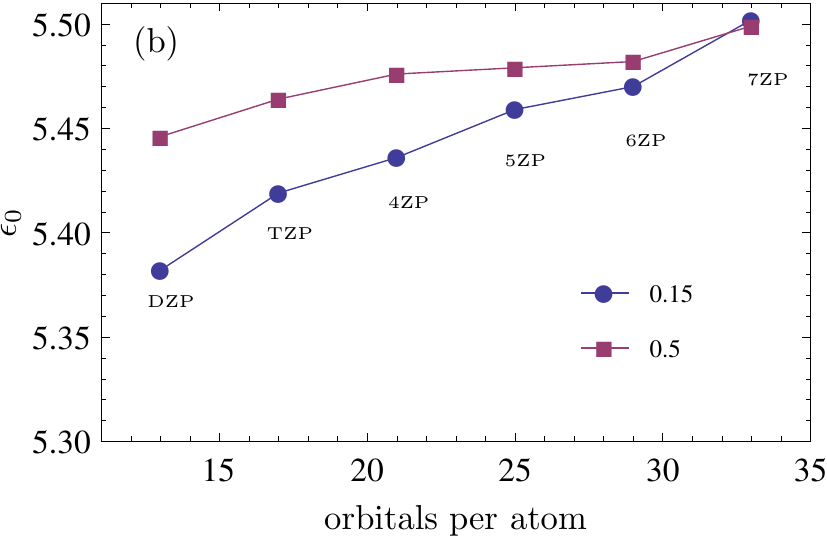}}
  \caption{
  \textbf{(a)}:
  Calculated macroscopic dielectric constant of diamond as a function
  of basis size, given in terms of orbitals per atom.
  The reference planewaves calculation is indicated by the horizontal red line.
  The datapoints cluster around two distinct curves: the upper curve
  corresponds to polarized basis sets, the lower curve to unpolarized basis sets.
  The number of $\zeta$ functions included is indicated by the labels SZ,
  DZ, TZ etc. The number of polarization orbitals for a given
  number of $\zeta$'s increases towards the right-hand side, as indicated
  for the case of the DZ basis. The energy shift is 10 meV and the split norm is 0.15.
  \textbf{(b)}:
  Calculated macroscopic dielectric constant of diamond as a function
  of basis size, for two different values of the split norm.
  \label{fig:c_static}}
  \end{figure}

Figure \ref{fig:c_static}(a) shows the calculated macroscopic dielectric
constant of diamond for a split norm of 0.15. 
The trend is similar to the case of silicon discussed in Sec.~\ref{sec.silicon}.
Also in this case basis sets without polarization orbitals lead to a slower
convergence rate as a function of basis size, and converge to a value 
significantly smaller than the reference planewaves result.
The converged value for the polarized basis set is $\eps_{0}=5.49$
for the 4Z4P basis, which includes 36 orbitals per atom. This value agrees very
well with corresponding planewaves result $\eps_{0}=5.47$.
As in the case of silicon a reasonably converged value ($\eps_{0}=5.41$, 
1\% smaller than the planewaves result) is already obtained using the TZP basis.

Figure \ref{fig:c_static}(b) shows the effect of the split norm on the
convergence of the macroscopic dielectric constant as a function of basis size.
In this case the trend is less clear than in Fig.~\ref{fig:si_static}(b),
however the same general conclusions apply: by increasing the split norm the dielectric
constant converges more rapidly and a plateau can be identified.

\subsubsection{Germanium}

\begin{figure}
  \begin{center}
  \resizebox{0.9\columnwidth}{!}{\includegraphics{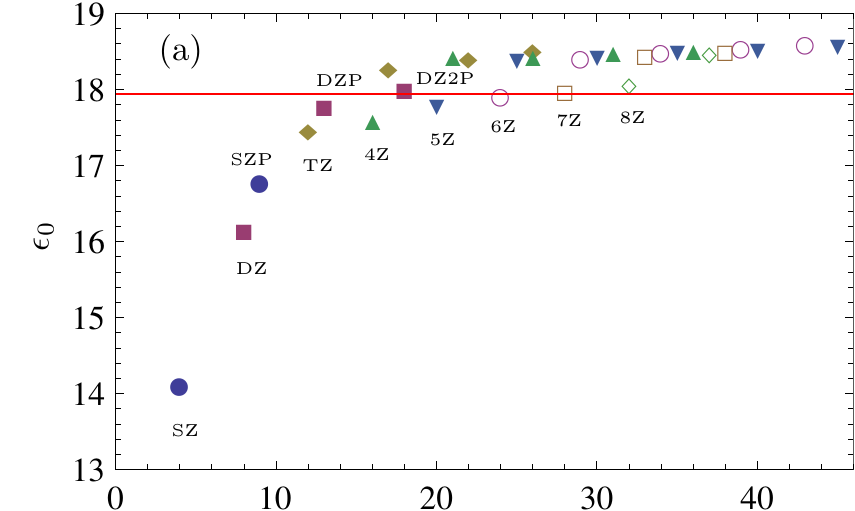}}
  \resizebox{0.9\columnwidth}{!}{\includegraphics{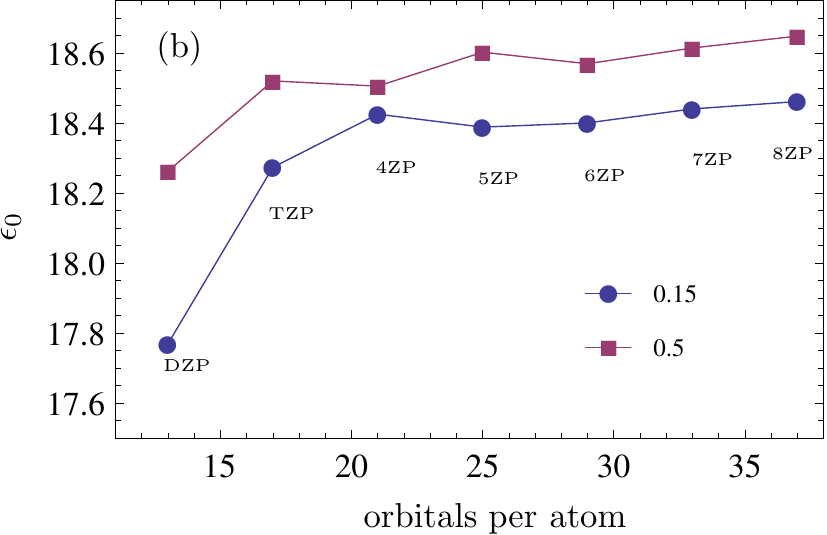}}
  \end{center}
  \caption{  
  \textbf{(a)}:
  Calculated macroscopic dielectric constant of germanium as a function
  of basis size, given in terms of orbitals per atom.
  The reference planewaves calculation is indicated by the horizontal red line.
  The datapoints cluster around two distinct curves: the upper curve
  corresponds to polarized basis sets, the lower curve to unpolarized basis sets.
  The number of $\zeta$ functions included is indicated by the labels SZ,
  DZ, TZ etc. The number of polarization orbitals for a given
  number of $\zeta$'s increases towards the right-hand side, as indicated
  for the case of the DZ basis. The energy shift is 10 meV and the split norm is 0.15.
  \textbf{(b)}:
  Calculated macroscopic dielectric constant of germanium as a function
  of basis size, for two different values of the split norm.
  \label{fig:ge_static}}
  \end{figure}

Figure \ref{fig:ge_static}(a) shows the calculated macroscopic dielectric
constant of germanium (split norm 0.15). 
Also in this case the basis sets with and without 
polarization orbitals appear to converge to different asymptotic values.
Similarly to the case of silicon and diamond the polarized basis sets
converge to a higher dielectric constant, $\eps_{0}=18.57$. This value
is 3\% larger than the reference planewave result of $\eps_{0}=17.94$.
Also in this case we observe that the TZP basis yields a dielectric
constant close to the fully converged value ($\eps_{0}=18.27$).

Figure \ref{fig:ge_static}(b) shows the effect of the split norm on the
convergence of the macroscopic dielectric constant as a function of basis size.
As in the other two cases, by increasing the split norm the calculated dielectric
constant converges more rapidly to its asymptotic value. 

\subsection{Frequency- and wavevector-dependent dielectric functions}\label{sec:frequency}

\begin{figure}
  \resizebox{0.9\columnwidth}{!}{\includegraphics{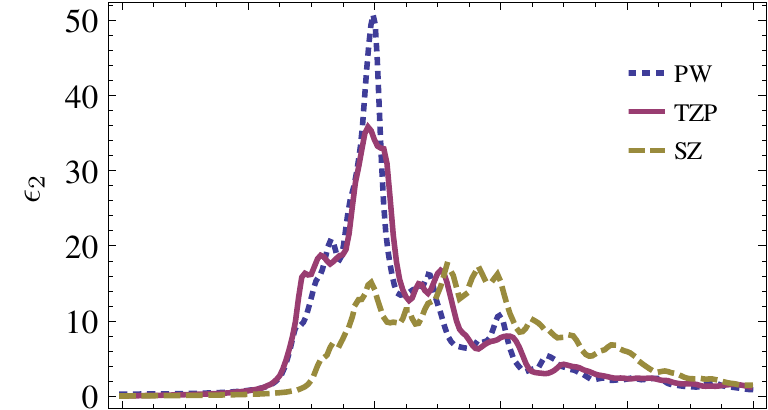}}
  \resizebox{0.9\columnwidth}{!}{\includegraphics{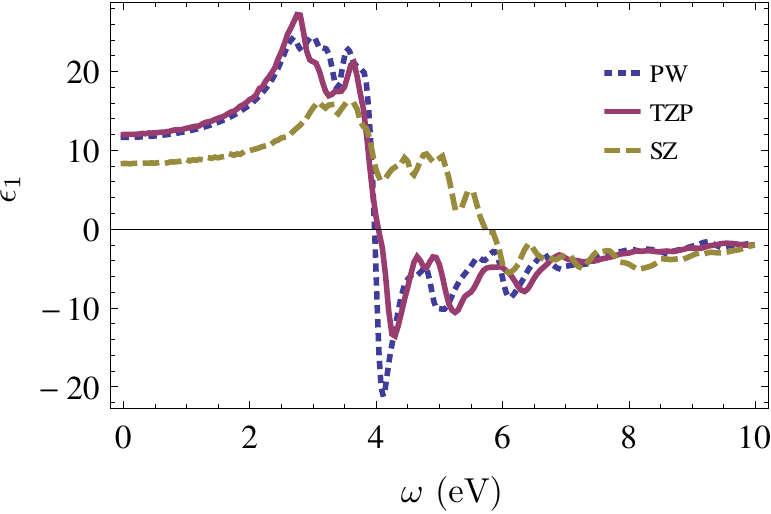}}
  \caption{  
  Calculated dielectric function of silicon: calculations using the SZ basis
  (dashed line), the TZP basis (solid line), and the reference planewaves  
  result (dotted line). A Gaussian smearing of width 0.1 eV is used.
  }
  \label{fig:si_wreal}
  \end{figure}

Figure \ref{fig:si_wreal} shows the frequency-dependent dielectric function
of silicon $\epsilon(\omega)$ for the minimal SZ basis set, the TZP basis
set, and the reference planewaves calculation.
The SZ basis performs very poorly, the spectral weight being incorrectly 
transferred from the main absorption peak to higher energy. This is consistent
with the small value of the macroscopic dielectric constant obtained with
the SZ basis in Fig.~\ref{fig:si_static}.

The TZP basis yields results in reasonable agreement with our reference planewaves 
result. The location of the main peaks and shoulders are correctly reproduced.
We note, however, some transfer of spectral weight from the main peak at $\sim$4 eV
to the shoulder at $\sim$3 eV, and a blueshift of the high-energy peaks.

\begin{figure}
  \resizebox{0.9\columnwidth}{!}{\includegraphics{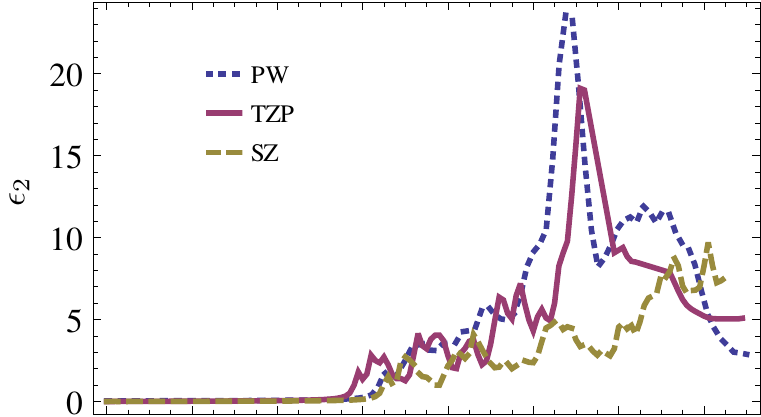}}
  \resizebox{0.9\columnwidth}{!}{\includegraphics{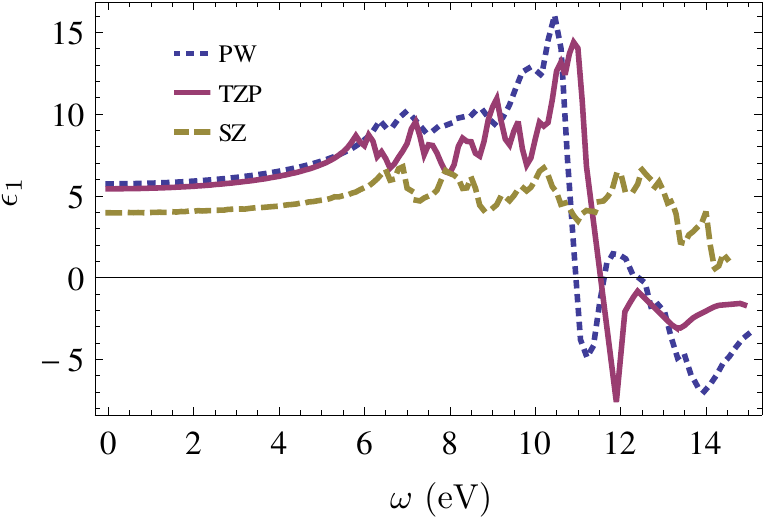}}
  \caption{
  Calculated dielectric function of diamond: calculations using the SZ basis
  (dashed line), the TZP basis (solid line), and the reference planewaves  
  result (dotted line). A Gaussian smearing of width 0.1 eV is used.
  }
  \label{fig:c_wreal}
  \end{figure}

\begin{figure}
  \resizebox{0.9\columnwidth}{!}{\includegraphics{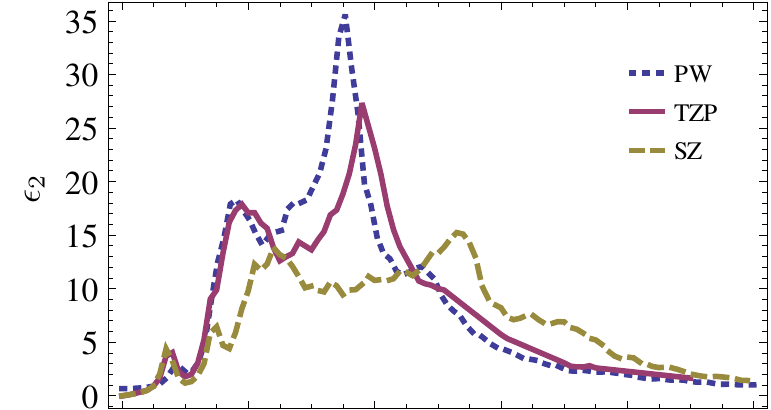}}
  \resizebox{0.9\columnwidth}{!}{\includegraphics{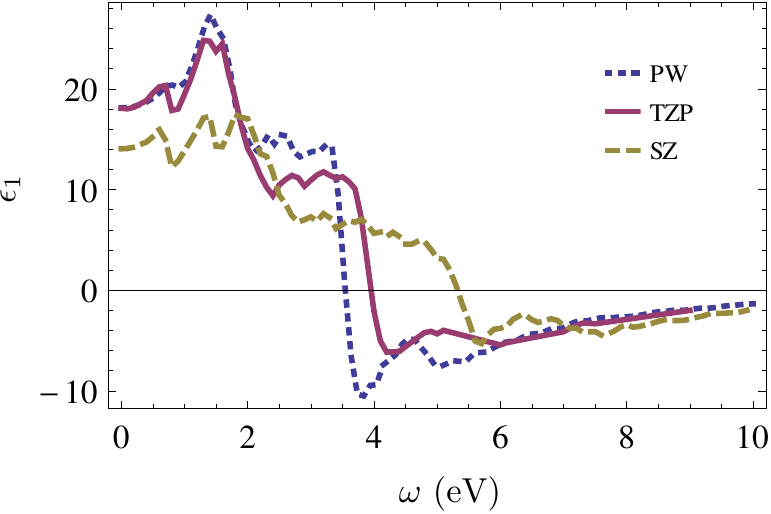}}
  \caption{
  Calculated dielectric function of germanium: calculations using the SZ basis
  (dashed line), the TZP basis (solid line), and the reference planewaves  
  result (dotted line). A Gaussian smearing of width 0.1 eV is used.
  }
  \label{fig:ge_wreal}
  \end{figure}

Figures \ref{fig:c_wreal} and \ref{fig:ge_wreal} show the frequency-dependent 
dielectric functions of diamond and germanium, respectively.
Also in these cases we compare the performance of the SZ basis and the TZP basis 
with the reference planewaves calculation. Conclusions similar to the case
of silicon can be drawn: the SZ basis misses the main peak and yields a 
blueshift of the other peaks, while the TZP basis is in better agreement
with the reference planewaves calculation.


\begin{figure}
  \resizebox{0.9\columnwidth}{!}{\includegraphics{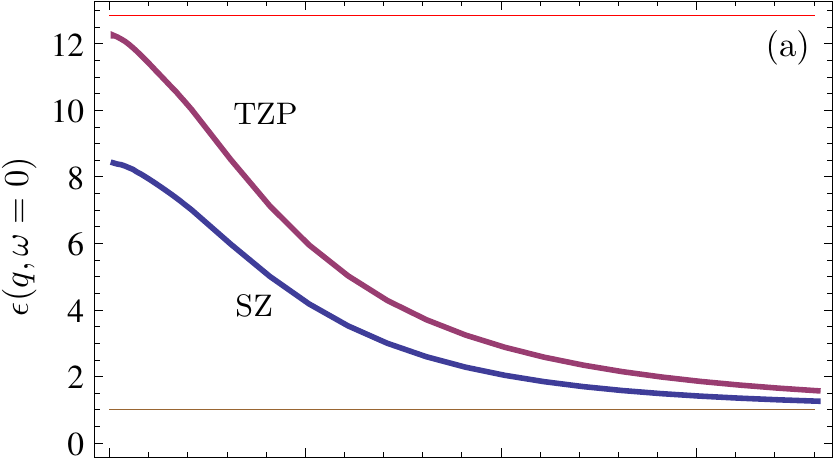}}
  \resizebox{0.9\columnwidth}{!}{\includegraphics{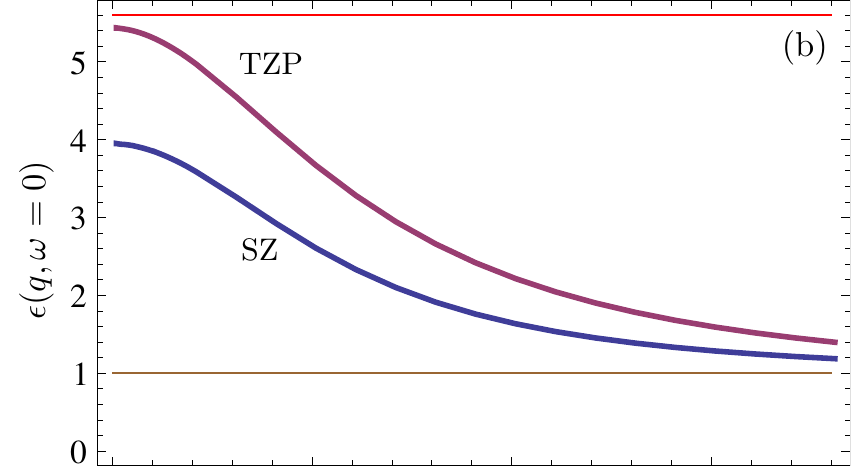}}
  \resizebox{0.9\columnwidth}{!}{\includegraphics{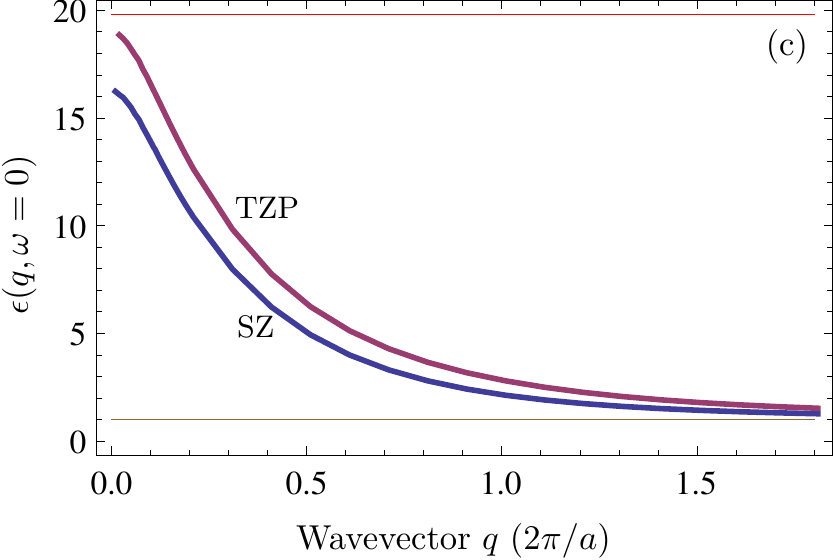}}
  \caption{
  Wavevector dependence of the dielectric function of (a) silicon, (b) diamond,
  and (c) germanium. We compare the performance of the SZ and TZP basis sets. 
  The upper horizontal line in each panel represents the static planewaves value
  $\eps_0$ and the lower horizontal line indicates the vacuum dielectric constant
  $\eps_{\rm vac} = 1$.
  \label{fig:q}}
  \end{figure}

Figure \ref{fig:q} shows the wavevector dependence of the dielectric function
$\epsilon(\qv,\omega=0)$ for silicon, diamond, and germanium, comparing
the performance of the SZ and the TZP basis sets. In all cases the wavevector
dependence shows the correct behavior,\cite{walter70} although the
SZ basis yields a smaller dielectric function across the full range of
wavevectors.

\section{Conclusions}\label{sec:conclusions}

We reported a systematic study of the performance of numerical pseudo-atomic 
orbital basis sets
of the {\tt SIESTA} code in the calculation of dielectric matrices 
in extended systems using the self-consistent Sternheimer approach 
of Refs.~\onlinecite{giustino10,huebener12}.
In order to cover a range of systems from more insulating to more metallic
character we presented results for the three semiconductors diamond, silicon, and 
germanium. 

Dielectric matrices, converged within the multi-$\zeta$ and polarization
scheme, fall within 3\% of reference planewaves calculations,
demonstrating that this method is promising.
We observed that the TZP basis already yields results very close to fully
converged values. This information may prove useful for practical calculations
of electronic excitations using pseudo-atomic orbital basis sets as in the
{\tt SIESTA} code. In particular the TZP basis yields the correct spectral
features in the long-wavelength frequency-dependent dielectric function.

We observed a consistent performance of the TZP basis across the three
systems considered, regardless of their more insulating (diamond) or metallic
(germanium) character. This may result from the different localization radii
which already account for the varying degree of localization of the density 
matrix in each material.

We also noted that polarization orbitals are critical 
for achieving good agreement with reference planewaves calculations.
This is somewhat expected since polarization orbitals precisely describe
the response to external fields.

We have investigated how the choice of the split norm influences the convergence
of the results. The increase of the split norm leads to multiple-$\zeta$ orbitals
with a wider distribution of localization radii and effectively improves
the completeness of the basis.

We point out that the localization radii of the basis sets discussed here
are rather large and therefore are not optimal for practical calculations.
Our choice was motivated by the need to systematically explore basis 
sets with many $\zeta$'s.
We expect that similar conclusions will be obtained by augmenting the basis
using diffuse numerical orbitals with similar localization radii as those
considered here.\cite{cofini07,ordejon09} The study of the performance of
diffuse orbitals deserves further investigation.

By providing a systematic assessment of the performance of pseudo-atomic
orbital basis sets including multiple-$\zeta$'s and polarization, 
the present work sets the ground for future studies of 
dielectric screening and electronic excitations in extended systems 
using local orbitals.

\subsection*{Acknowledgments}
The authors would like to thank E. Artacho for fruitful discussions. This work
is funded by the European Research Council under the European 
Community's Seventh Framework Programme, Grant No. 239578, and Spanish MICINN
Grant FIS2009-12721-C04-01.

\appendix
\section{Bloch phase factors in the variation of the density matrix
for periodic systems}\label{app:qR}

The evaluation of the variation $\Delta n_{[\qv,\rv,\omega]}$ of the density matrix 
using Eq.~(\ref{eq:dn_matrix}) requires the introduction of the Bloch phase
factors ${\rm exp}(i\kv\cdot\Rv)$, ${\rm exp}(i\qv\cdot\Rv)$, and ${\rm exp}(i\qv\cdot\rv)$
at various stages. We proceed as follows.

First we merge into a single index $I$ the basis index $i$ and the unit
cell vector $\Rv_I$ of the cell for each orbital $\phi_i(\rv'-\Rv_I)$: 
$I=(i,\Rv_I)$ and $\phi_I(\rv')=\phi_i(\rv'-\Rv_I)$.
Using this notation we rewrite Eq.~(\ref{eq:dn_matrix}) as follows:
\begin{eqnarray}\nonumber
  & & \hspace{-1.0cm}\Delta n_{[\qv,\rv,\omega]}(\rv') =
  \frac{2}{N_{\kv}}\sum_{v\kv\sigma=\pm}\sum_{IJ}c^*_{vi\kv}
  \Delta c^\sigma_{vj\kv[\qv,\rv,\omega]} \\
  &  &\hspace{-.8cm} \times e^{-i\kv\cd(\Rv_I-\Rv_J)}e^{i\qv\cd\Rv_J}
  e^{-i\qv\cd\rv'}\phi_I(\rv')\phi_J(\rv'),\label{eq.app.1}
  \end{eqnarray}
where $i$ in $c^*_{vi\kv}$ is still the orbital component of the composite indices $I=(i,\Rv_I)$,
and similarly for $j$.
In order to evaluate Eq.~(\ref{eq.app.1}) we first calculate the matrix
  \begin{equation}
  \Delta n^{(1)}_{ij\kv[\qv,\rv,\omega]}
   = 2\sum_{v\sigma=\pm} c^*_{vi\kv}
  \Delta c^\sigma_{vj\kv[\qv,\rv,\omega]}.
  \end{equation}
Second, we perform the sum over the wavevectors $\kv$ and introduce the
phase factors ${\rm exp}[i\kv\cdot(\Rv_I-\Rv_J)]$:
  \begin{equation}\label{eq.app.2}
  \Delta n^{(2)}_{IJ[\qv,\rv,\omega]} = 
   \frac{1}{N_{\kv}} \sum_{\kv} \Delta n^{(1)}_{ij\kv[\qv,\rv,\omega]}
   e^{-i\kv\cd(\Rv_I-\Rv_J)}.
  \end{equation}
Third, we include the phase factor ${\rm exp}(i\qv\cdot\Rv_J)$:
  \begin{equation}\label{eq.app.3}
  \Delta n^{(3)}_{[\qv,\rv,\omega]}(\rv') = 
  \sum_{IJ} \Delta n^{(2)}_{IJ[\qv,\rv,\omega]}
   e^{i\qv\cd\Rv_J} \phi_I(\rv')\phi_J(\rv'),
  \end{equation}
and finally we introduce the factor ${\rm exp}(-i\qv\cdot\rv)$:
  \begin{equation}\label{eq.app.4}
  \Delta n_{[\qv,\rv,\omega]}(\rv') =
  e^{-i\qv\cd\rv} \Delta n^{(3)}_{[\qv,\rv,\omega]}(\rv').
  \end{equation}
This final phase factor is added only after the real space density response 
has been evaluated on the grid.

The reason for proceeding as described here becomes evident if we make
the observation that we only need to calculate the
density variation inside the fundamental unit cell.
This implies that we only need to work with basis orbitals 
belonging to the fundamental unit cell or which are nonvanishing
in this cell. As a consequence we can calculate Eq.~(\ref{eq.app.2}) only
for those $I,J$ which lead to finite overlap with the fundamental cell.
Furthermore, it is convenient to calculate $\Delta n^{(2)}_{IJ[\qv,\rv,\omega]}$
only for the index $I$ belonging to the fundamental
unit cell [i.e.\ $I=(i,0)$] and the index $J$ over the orbitals with
finite overlap with this cell: $J=(j',\Rv_J)$. This allows us to rewrite
Eq.~(\ref{eq.app.2}) as:
  \begin{equation}\label{eq.app.2b}
  \Delta n^{(2)}_{ij'[\qv,\rv,\omega]} = 
   \frac{1}{N_{\kv}} \sum_{\kv} \Delta n^{(1)}_{ij\kv[\qv,\rv,\omega]}
   e^{-i\kv\cd\Rv_{ij'}},
  \end{equation}
where $\Rv_{ij'}=\Rv_{I}-\Rv_{J}$ is the vector pointing from 
orbital $I$ to $J$ and $I$ is still $I=(i,0)$.
The matrix $\Delta n^{(2)}_{ij'[\qv,\rv,\omega]}$ in Eq.~(\ref{eq.app.2b}) 
has the first dimension equal to the number of orbitals in the unit cell, 
and the second dimension equal to the number of orbitals that are non-zero 
in the fundamental unit cell. This is a sparse matrix 
and it is stored using the sparse matrix representation of {\tt SIESTA}. 

The reason for evaluating Eqs.~(\ref{eq.app.3}) and (\ref{eq.app.4})
separately is that the phase factor ${\rm exp}(i\qv\cd\Rv_J)$ cannot be added 
in the same way as the $\kv$-dependent factor into Eq.~(\ref{eq.app.2b}),
since it depends 
on the absolute position of the cell $\Rv_J$ and not on the relative
position of the orbitals $\Rv_J-\Rv_I$. Using this alternative notation
we can  rewrite Eq.~(\ref{eq.app.3}) as:
  \begin{equation}\label{eq.app.3b}
  \Delta n^{(3)}_{[\qv,\rv,\omega]}(\rv') =
  \sum_{i'j'} \Delta n^{(2)}_{ij'[\qv,\rv,\omega]}
   e^{i\qv\cd\Rv_{j'}} \phi_{i'}(\rv')\phi_{j'}(\rv'),
  \end{equation}
where $i',j'$ refer to orbitals which are non-zero in the unit cell,
and $i$ is the replica of $i'$ belonging to the fundamental unit cell.

Our procedure allows us to use the the sparse matrix $\Delta n^{(2)}_{ij'[\qv,\rv,\omega]}$
as the working quantity for the self-consistent cycle (i.e.\ for 
charge-density mixing and convergence tests).
The scheme outlined here uses the sparse matrix representation built in {\tt
SIESTA} 
and requires only small changes to existing subroutines that manage the evaluation 
on the real space grid. It can therefore easily make use of subroutines 
that manage the density matrix during the scf procedure, such as density-mixing. 
Furthermore, these steps ensure that the computational overhead of our procedure
is minimal.

\end{document}